\documentclass{aa}
\usepackage{graphicx}

\def\es{erg~s$^{-1}$}

\begin{document}

\title{The detection of diffuse emission in HCG 16 with XMM-Newton}

\author{E. Belsole\inst{1} \and J-L. Sauvageot\inst{1} \and  T.J. Ponman\inst{2}\and H. Bourdin\inst{3}}
\offprints{E. Belsole email:ebelsole@cea.fr}

\institute{Service d'Astrophysique, CEA Saclay, L'Orme des Merisiers 
B\^at 709., F-91191 Gif-sur-Yvette Cedex, France\\
\email{ebelsole@cea.fr}\\
\email{jsauvageot@cea.fr}
\and 
School of Physics and Astronomy, University of Birmingham. Edgbaston, Birmingham B15 2TT.\\
\email{tjp@star.sr.bham.ac.uk}
\and 
Observatoire de la C\^ote d'Azur, B.P. 4229, F-06304 Nice Cedex 4, France.\\
\email{bourdin@obs-nice.fr}
}

\date{Received January 28, 2002; accepted October 10, 2002}

\abstract{We report results obtained from analysis of the XMM-Newton observation of the compact group of galaxies HCG 16. It is a peculiar system composed of 7 spirals, 6 of which are active, and its nature as a bound system has been much debated. The EPIC camera observations give new insights into the X-ray parameters describing the physical status of the group. We detect  diffuse X-ray emission with  a rather elliptical morphology which extends to at least a radius of 135 $h^{-1}_{50}$ kpc from the group centre. The spectrum within this region is well modelled by a thermal plasma with a temperature of 0.49$\pm$ 0.17 keV, and a non-zero metallicity. We measure a bolometric X-ray luminosity of 9.6$\times$10$^{40}$ $h^{-2}_{50}$ \es which may be only a small fraction of the total luminosity because of the limit in spatial detection arising from the high background level. Despite  its  low temperature and luminosity, HCG 16 obeys the $L_X-T$ relation obtained for brighter galaxy groups even if it lies in a very extreme position. The properties of the diffuse emission confirm the bound nature of HCG 16 even if the gas trapped in the potential well may not yet be virialized. This reopens the debate about the real nature of spiral-dominated galaxy groups, and on their role in a more general cosmological context.
}
\maketitle
%
\section{Introduction}
Small groups of galaxies are the most common configurations of
aggregated systems in the Universe (Tully 1987 and references therein)
and they continue to be the subject of intense study. Despite much 
attention in the last decade, there is continued debate as to whether
these systems are genuinely gravitationally bound (Hickson \& Rood 1988) 
or if they are occasional alignments along the line of
sight belonging to more extended systems (Mamon 1986), clusters (Walke
\& Mamon 1989), or cosmological filaments (Hernquist et al. 1995).  In
particular the interest in compact groups of galaxies, of which the
most widely studied catalogue was compiled by Hickson (1982), revealed a
series of peculiarities as compared to galaxies in clusters or to
isolated galaxies of the same type (see Hickson 1997 and Mulchaey
2000 for reviews).

Galaxy groups have been observed in the X-ray since the advent of the
{\em Einstein} satellite. In fact, the detection of hot gas trapped in
the group potential well represents  strong proof of the bound nature
of these systems.

\noindent Several works based on ROSAT/PSPC and ASCA data revealed a
variety of morphological and dynamical behaviour in galaxy groups
(Pildis et al 1995; Saracco \& Ciliegi 1995,; Ponman
et al. 1996 (hereafter PBEB), Heldson \& Ponman 2000b, Mulchaey \& Zabludoff 1998).

More  than 50\%  of the 32 ROSAT/PSPC-observed compact galaxy groups (pointed observations) analysed by  PBEB exhibit diffuse X-ray emission with temperatures lower than 1.5 keV: the authors found that correcting for the selection effects, the fraction of HCGs with $L_X$(bolometric) $> 10^{41.1} h_{50}^{-2}$ \es ~was 75\%.  This high fraction ruled out the hypothesis that groups are aligned configurations within cosmological filaments. 

Galaxy groups show flatter surface brightness profiles and
steeper ``luminosity-temperature'' ($L_X-T$) and ``velocity dispersion-temperature'' ($\sigma-T$) relations compared
to clusters. This behaviour becomes even more noticeable for very-low
temperature (kT $<$ 1 keV) systems (Helsdon \& Ponman 2000a; Mulchaey 2000, Xue \&
Wu 2000) where the departures from the self-similar scaling laws traced
for galaxy clusters have been explained by models of pre-heating of the
gas before it falls into the group (Cavaliere, Menci \& Tozzi 1997;
Ponman et al. 1999; Balogh, Babul \& Patton 1999; Dos Santos \& Dor\'e 2002). 

These low
temperature systems are puzzling for other reasons. Whereas
it is well established that a gravitationally bound intergalactic
medium is a common property of spiral-poor groups, the results are
still controversial for groups composed of more than 50\% of spiral
galaxies (Pildis et al 1995; Saracco \& Ciliegi 1995; Mulchaey et al. 1996), which accounts for the majority of groups.

A rare opportunity to investigate these questions is presented by an
extreme example of a very nearby (z = 0.013) compact group: the Hickson Compact Group 16
(hereafter HCG 16). This group is peculiar because it is composed of 7
galaxies (4 originally detected by Hickson (1982), with 3 more added
in a circle of 15\arcmin~ by Riberio et al. (1996) on the basis of
radial velocity measurements), of which {\em all} are
spiral and 6 of the 7 are active.  It is one of the densest groups in
the Hickson catalogue, with a mean velocity dispersion of $\sim$ 100 km
s$^{-1}$ and a median projected galaxy-galaxy  separation of 88 $h_{50}^{-1}$ kpc (Hickson et al. 1992)\footnote{throughout this paper we assume $H_0$ = 50 km s$^{-1}$ Mpc$^{-1}$; $h_0$ is the Hubble constant in units of 50 km s$^{-1}$ Mpc$^{-1}$}.

Bachall et al. (1984) observed HCG 16 with {\em Einstein}, determining an
X-ray luminosity of 2$(\pm$1) 10$^{41}$ $h_{50}^{-2}$ \es, but with the spatial resolution
of the instrument the authors were unable to separate the emission of
the galaxies from a more diffuse emission.

Using the ROSAT/PSPC, HCG 16 was later studied by Saracco \& Ciliegi (1995), who suggested
that the X-ray emission was mainly due to the galaxies. However PBEB and
Dos Santos \& Mamon (1999, hereafter DSM99), analysing the same ROSAT/PSPC 
data set,
found diffuse X-ray emission far away ($\sim$200 kpc) from the
galaxies and with a temperature of the order of 0.3 keV, which makes
HCG 16 the coolest compact group yet detected.

The detailed spatial analysis of DSM99 ruled out
the possibility that the hot gas is virialized. They further suggested
that the contribution  to the gas enrichment from galactic winds could
be an important test to understand the dynamics and
evolutionary history of this group.

In this work, we present a detailed analysis of the XMM-Newton
observation of the HCG 16, focusing on detection of the hot X-ray gas
(we do not discuss the galaxy emission, see Turner et al. 2001a). At the redshift of the group, 1\arcmin~ is equivalent to 22.5 kpc.

\section{Observations and data reduction}

\subsection{Observations}
The XMM/EPIC (2 EMOS and 1 EPN camera - Turner et al. 2001b, Holland
et al. 1996, Str\"uder et al. 2001) first light observation was a 65
ks exposure of HCG 16. This consisted of two separate exposures (only 1 for EPN) for which each camera was turned on individually (there was no standard mode). Information on the roll angle is not available for these very early data, but we were able to superimpose the two MOS camera image obtained in raw data, and no evident roll angle variation is observable. The observation log is listed in Table 1.

\begin{table*}[!ht]
\begin{center}
\caption{The observation log. The net exposure is the sum of exposure 1 and 2 and after flare rejection.}
\begin{tabular}{|c|c|c|c|c|}
\hline
\hline
	& EMOS 1  & EMOS 2 & EPN \\ 
\hline
exposure 1 (s) &  55203	& 48302	&  32399  	\\	
comments	& 50\% flare rejection & CCDs 2 \& 5 noisy	& many bright pixels	\\
		&		& 50\% flare rejection & 40\% flare rejection \\
\hline
exposure 2 (s) &  16910	& 17137	&     0 \\
comments	&  --- 	& CCD2 off; CCD5 noisy	&  ---	\\
\hline
net exposure (ks) & 45905	& 43858	&     12525	\\
\hline
useful data  (ks) & 45905	& 43858	&     0	\\
\hline
\end{tabular}
\end{center}
\end{table*}

A first data processing has been done with a preliminary version of
the SAS (Science Analysis Software) pipeline scripts EMCHAIN (EMOS)
and EPCHAIN (EPN). Because of the early stage of the experiment the
data files are not in standard configuration, and thus we are unable to
process the data with the current (5.3 version) of the SAS
software. Consequently the data are treated using software developed
in Saclay for calibration purposes. The output products are consistent
with the SAS for downstream compatibility. The processing takes into account 
the electronic noise cleaning and the difference in exposure time per
CCD. This data pre-processing eliminates most of the low energy noise,
but from a conservative point of view we limited all analysis to
energies above 0.2 keV. A higher noise is detected in some CCDs (CCD2 and 5 for EMOS 2 and CCD 4 for EMOS 1), likely due to the relatively high temperature of some CCDs at this stage of the
experiment (they were later cooled). For example, the threshold for telemetry transmission was higher for the CCD2 of MOS2, representing an additional source of noise. Furthermore, this CCD was turned off for one of the EMOS 2 exposures, decreasing the quality of the data from this spatial region.

High background time intervals (due to solar flares) were rejected,
for the EMOS cameras, by a careful examination of the 10 to 12 keV
light curve. The drop of the EMOS effective area at high energy allows
us to be sure that the light curve variation is not due to source
variability in this energy band. Events were grouped in 100 s bins
and all bins with more than 18 counts were rejected.  After this
filtering, the useful exposure time is 45 ks.

\noindent The EPN observation was made with the very first on-board instrument
configuration, which is rather different from that used currently, and
which has not been yet calibrated. Furthermore, the available blank-sky
 background observations were obtained in the standard configuration
setting. Because of these problems, we have not attempted to use the PN
data for our analysis, restricting ourselves to the 45 ks of useful 
EMOS data. 

\begin{figure}[h]
\vskip 0.5cm
\begin{center}
\resizebox{8.8cm}{!}{\includegraphics{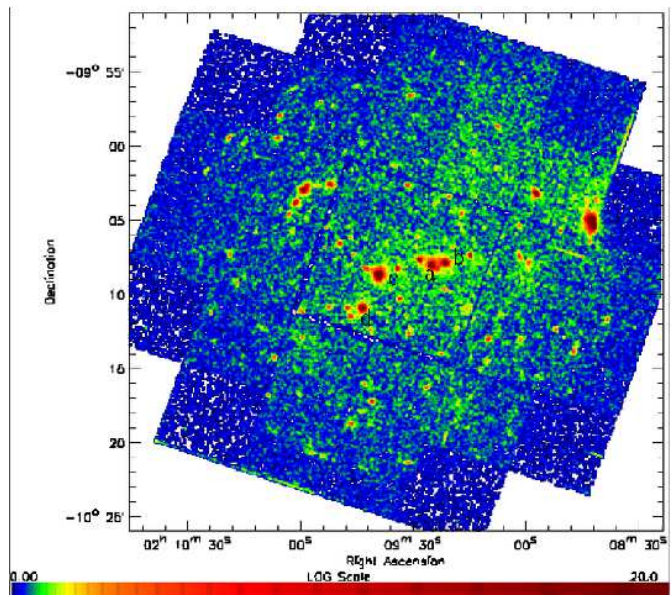}}
\caption{HCG 16 EMOS 1+EMOS 2 image in the energy band 0.2-7.0 keV. The 4 main galaxies are labelled as in Hickson (1982).}
\label{fig:fig1}
\end{center}
\end{figure}

Figure \ref{fig:fig1} shows the EMOS 1 and EMOS 2 summed count rate image of the HCG 16 pointing in the energy band [0.2 -7.0] keV, smoothed with a Gaussian filter of $\sigma$ = 12\arcsec. The image is not corrected for vignetting. 
The four main galaxies are marked as in Hickson (1982). 

\subsection{Background estimate}

The estimate of the background level is a crucial point since we are
interested in a very low surface-brightness, diffuse source. Because of the
high background level we can assume that the contribution of any diffuse gas
emission at large radii from the group centre is negligible. This
allows us to make an estimate of the background in the same field
(i.e. a local background). We consider that the XMM-Newton background
is composed of two main  components:

\begin{enumerate}
\item an astrophysical X-ray background which is the combination of a soft (E$<$ 1.5 keV) component, mainly due to the local bubble, and a hard component  due to unresolved cosmological sources (mainly AGN).

\item an induced cosmic ray background which dominates at high energy
($>$ 5 keV) and induces fluorescence lines (Al, Si, Cu, Au) from the shielding
of the camera and the detector itself.
\end{enumerate}

The two components deposit an indistinguishable signal on the
detector, but have to be considered as completely different in the
data analysis.  The first component represents the X-ray photons that
are collected by the telescope and focused on the camera, and so they are
affected by vignetting. Conversely, the 2nd component is not affected
by vignetting because the particles pass throughout the instrument as a
whole. This means that to optimise the analysis of the X-ray
background it is better to estimate the particle background before the
application of the vignetting correction (see Belsole et al. 2001 for
details).

EMOS detections outside the field of view (FoV) of the telescope are only due to particle events. One observation is not sufficient accurately to estimate the particle background, due to the small number of detected events. We thus summed the performance verification (PV) phase EMOS observations to achieve a total exposure time of $2 \times 10^6$ s, and considered the out-of-FoV events. These events are detected by all but the central CCD, which has no out-of-FoV region. The spatial and spectral distribution of this particle emission was extended into the FoV, for each CCD, using a Monte Carlo simulation under the following assumptions:

\begin{itemize}
\item  the particle spectrum outside and inside the FoV are identical:

We use the $2 \times 10^6$ s PV observation. The only spectrally useful region to compare directly the particle spectrum inside and outside the FoV is above 10 kev. In this spectral range, the region inside the FoV is essentially source-free due to the drop in effective area. For each CCD we verified that the spectrum above 10 keV is the same for the outside and inside FoV regions, except for a normalisation factor (of the order of 10\%) due to the shielding of the telescope. It is thus an acceptable hypothesis to assume the equivalence between the continuum spectrum inside and outside the FoV in the whole energy band of EPIC.

\item  it has a uniform spatial  and spectral distribution in each CCD:

We  looked for variations between CCDs in the same energy band above 10 keV. Examination of the spectrum of each CCD shows that there is some variation corresponding to the fluorescence lines of gold (11.47 keV, present only in CCD2 and 7 for EMOS 1, for example), but that the spectrum of the continuum is very similar from CCD to CCD, only showing a variation of about 5\%. We thus estimate the particle spectrum and spatial distribution of the central CCD (which has no out of FoV regions) from the median between CCD 3 and 6, which are closest to it following the camera geometry. Using this approach, we are able to accurately reproduce the high energy ($>$10 keV) spatial variations due to gold fluorescence lines because at these energies there are no source contributions. We note however, that the Al (1.48 keV) and the Si (1.74 keV) fluorescence lines are not spatially constant. Reproduction of the distribution of these lines is severely complicated by source contributions, and will require careful analysis of CLOSED{\footnote{Observations obtained during the calibration phase using an aluminum filter of 1mm thickness, which stops the X-ray photons but not the particles.}} data, a statistically useful quantity of which has only recently been released. We take a conservative approach because of the relative weakness of our source, and exclude  data from 1.4 to 2.0 keV (fluorescence Al and Si lines) in all subsequent analysis.

\item  the spectrum of the particle background does not change significantly in time. 

While the average induced cosmic ray background level  has been shown to be constant within $\pm 10\%$ (see e.g., Arnaud et al. 2001), once periods affected by solar flares are excluded, no significant spectral variation  has yet been  detected.
\end{itemize}

\noindent In conclusion, the continuum spectrum of the particle background inside the FoV can be well represented using the out of FoV distribution, extended into the FOV using the Monte Carlo method under the assumptions listed above.

EMOS 1 and EMOS 2 particle event lists, with a total simulated
exposure time of 10$^6$ s, were generated in the same format as
real EMOS observations and were used for the subsequent
image and spectral analysis. As the analysis of the particle
distribution was done on the out of FoV events and the flux in this
region is lower than in the FoV because of the shielding of the
camera, the particle count rate and the source count rate were
normalised in the 10 to 12 keV energy band.

\section{Analysis and Results}
\subsection{Image and preliminary spectrum}

As a first approach in detecting any diffuse X-ray emission, we
smoothed the 0.3 to 7 keV EMOS 1 image to find the spatial region where
the group diffuse signal is mainly located. To optimise the S/N of the
image we performed a preliminary spectral analysis by extracting a
spectrum in a circle of 6\arcmin ~in radius around the optical centre
of the group (i.e. where the smoothed image shows a significantly denser
emission), excluding the signal from the galaxies. We subtracted from
it the spectrum of the 10\arcmin~- 12\arcmin~ ring in order to have an estimate
of the local background. This rough analysis allows us to calculate an upper limit to the flux and to determine that the peak of the X-ray diffuse emission is around 1 keV
(as previously shown in ROSAT data). Also, no emission is detected
above 2 keV where the spectrum is dominated by the emission from
individual galaxies (see Turner et al 2001a). Because this work is
focused on the group diffuse X-ray emission, unless otherwise stated
we limit the following analysis to the soft energy band [0.2-1.4] keV, the upper energy limit being determined by the contamination of the Al and Si fluorescence lines (see Sect. 2.2).

\subsection{Wavelet analysis} 
The wavelet de-noising is a multi-scale image reconstruction tool. The noise contribution to the signal is removed by selecting its corresponding coefficients in the wavelet space. 
A new iterative wavelet de-noising algorithm (based on the {\em \`a trou} algorithm) especially suited for images with few counts is used here. Within this algorithm, thresholds are computed analytically for a Poisson noise distribution (see Bourdin et al. 2001).
This algorithm has been written to be used for restoring the surface brightness of the X-ray emitting intra-cluster gas which is supposed to have no steep discontinuities. In the case of a group like HCG 16, the individual galaxy emission introduces a high level of discontinuity with a dynamic of three orders of magnitude. A crude application of our algorithm to such data would be good at restoring the emission of the extended gas but the restored image would be contaminated by artifacts around these highly emitting sources.

In order to exclude this contribution, a count threshold was first defined on the image, and the bright sources identified. Then the flux inside the source emitting regions, as well as in the CCD gaps, was interpolated so as to create a new image without steep discontinuities. The image was de-noised with the restoration algorithm and the sources were then re-added. 

We have defined a counts (Poisson's statistic, i.e. no vignetting
correction is done) image of 4\farcs.1 pixel size where the contributions of
the two EMOS cameras are added. The energy band 0.2-1.4 keV has been
chosen so as to be sure that the detection excludes the Al and Si fluorescence
lines. An image in the same energy band is also obtained for each camera, from the particle event lists. The EMOS particle images are then coadded. 

The HCG16 counts image is de-noised with the previously described wavelet de-noising  procedure. 
 The threshold probability adopted for the HCG 16 image is of $10^{-6}$, meaning that there is a probability of $10^{-6}$ that the detected structure comes from noise, or in other terms, the structures detected in the image are significant at 4.5 $\sigma$. The adopted threshold is quite severe, but due to the low S/N we prefer to adopt such  a conservative approach. We detect significant extended emission only at scale 6, corresponding to 4\farcm.4 at the pixel size of the image.  We do not detect any extended emission at higher frequency scales, indicating that if more clumpy structures exist, these cannot be significantly detected with these data. We are unable to distinguish between the real galaxy extension, the effect of the Point Spread Function (PSF) wings and the actual diffuse emission. However, it is unlikely that at the scale of detection the signal comes from the galaxies only (see Sect. 3.4.2).
\begin{figure*}[!]
\begin{center}
\resizebox{13.0cm}{!}{\includegraphics{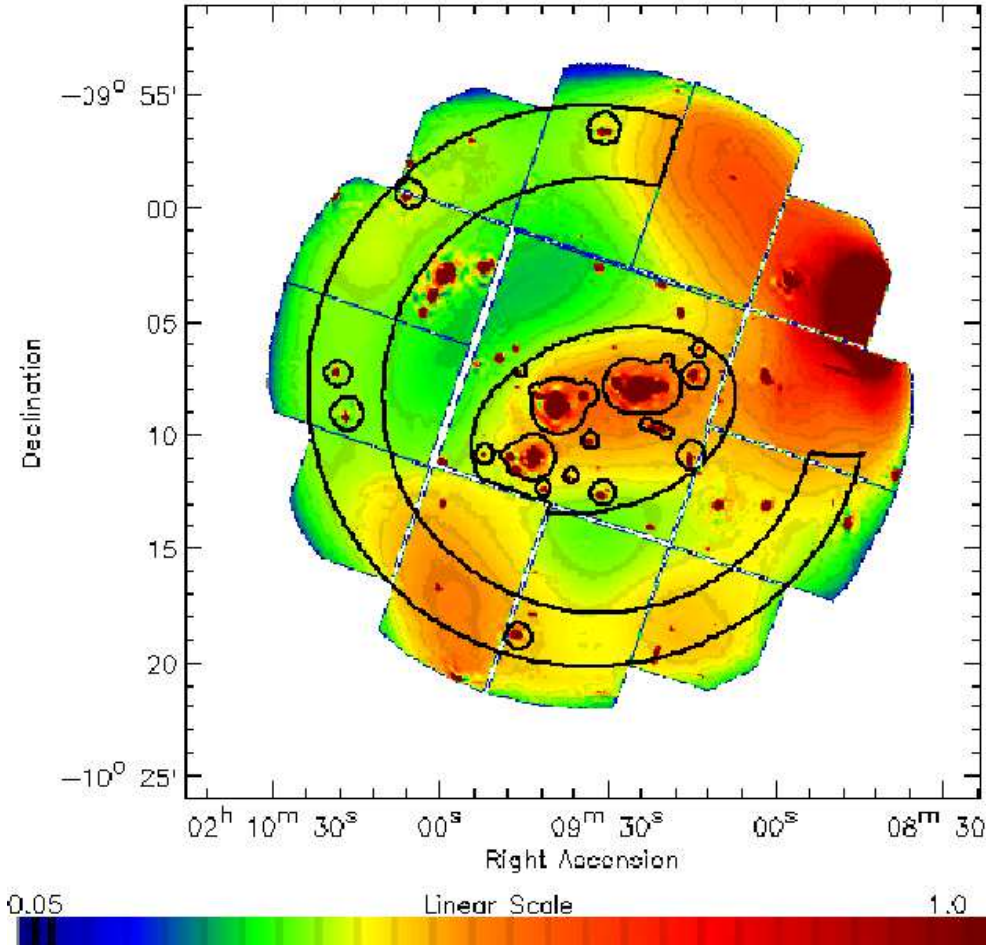}}
\caption{ EMOS 1+EMOS 2 wavelet image  in the energy band 0.2-1.4 keV. The ellipse at the centre of the figure denotes the region used for spectrum extraction. The ring from which the background spectrum was extracted is also shown to indicate the distance from the elliptical region. A slightly different ring is used for EMOS 1 and EMOS 2 in order to exclude the noisy CCDs.}
\label{fig:fig2}
\end{center}
\end{figure*}

The particle background image is smoothed with a Gaussian kernel of $\sigma = 1\farcm.4$, comparable to the scale of the large structures in the wavelet image. This operation allows us to take into account boundary effects which would be ignored if we just subtracted a constant value per CCD.
The smoothed particle image was then normalised by the exposure time and subtracted from the denoised counts image. We then corrected the image for vignetting and the net result  is shown in Fig.~\ref{fig:fig2}. To aid in the interpretation of the observed features, we superimpose on this figure the CCD gaps.

The wavelet detection of diffuse gas between galaxies exhibits an elliptical shape which extends up to a radius of $\sim$6\arcmin ~($\sim$ 135 kpc) from the optical centre of the group to the NW-SE direction and to 4\farcm.5 in the other direction. 

However, there are some  noticeable effects in the image (we show in Fig. \ref{fig:ccdposition} the position of CCDs in the HCG16 sky coordinates):

\begin{figure}[!ht]
\centering
\includegraphics[scale=0.20,angle=0,keepaspectratio]{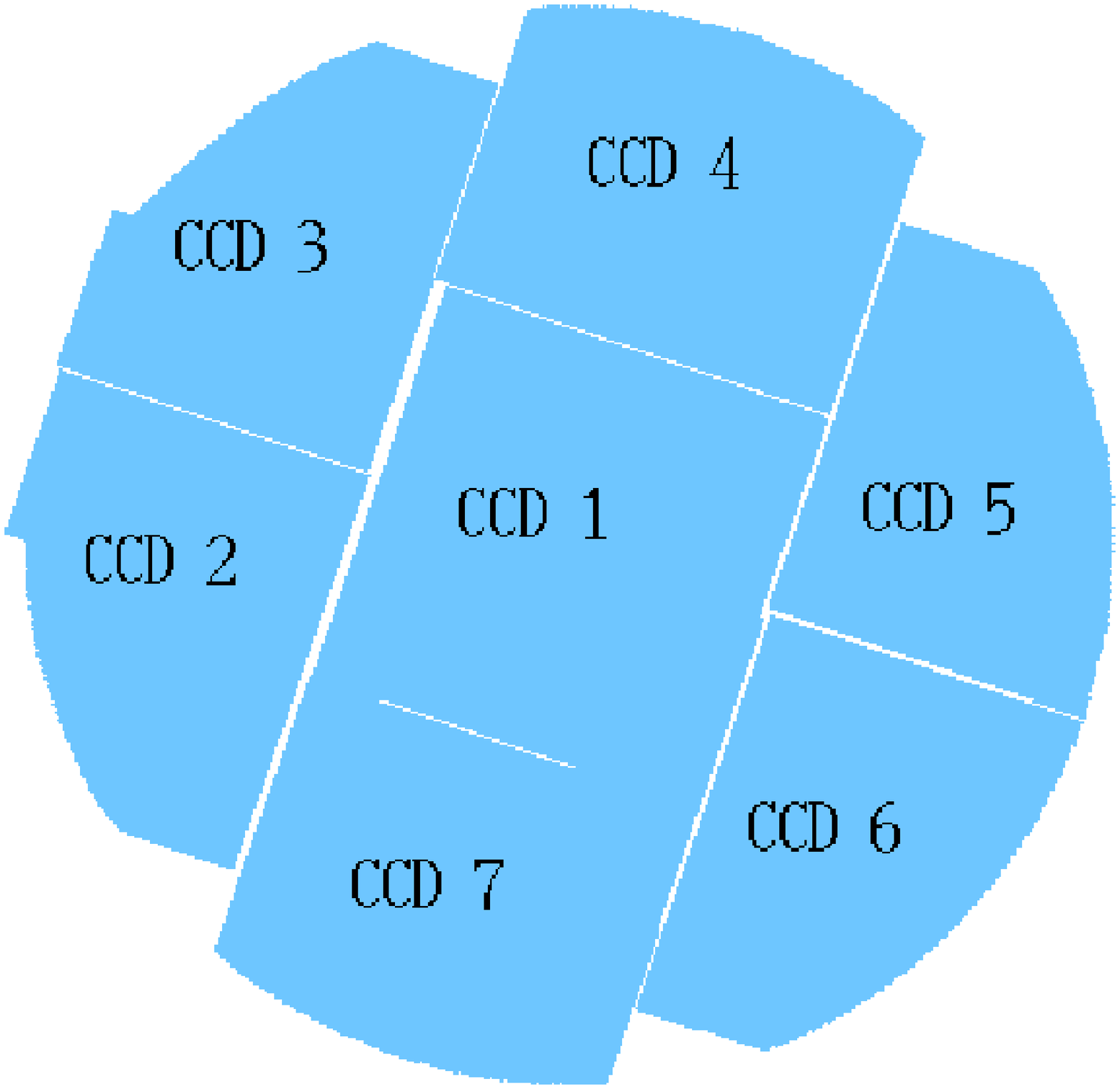}
\includegraphics[scale=0.20,angle=0,keepaspectratio]{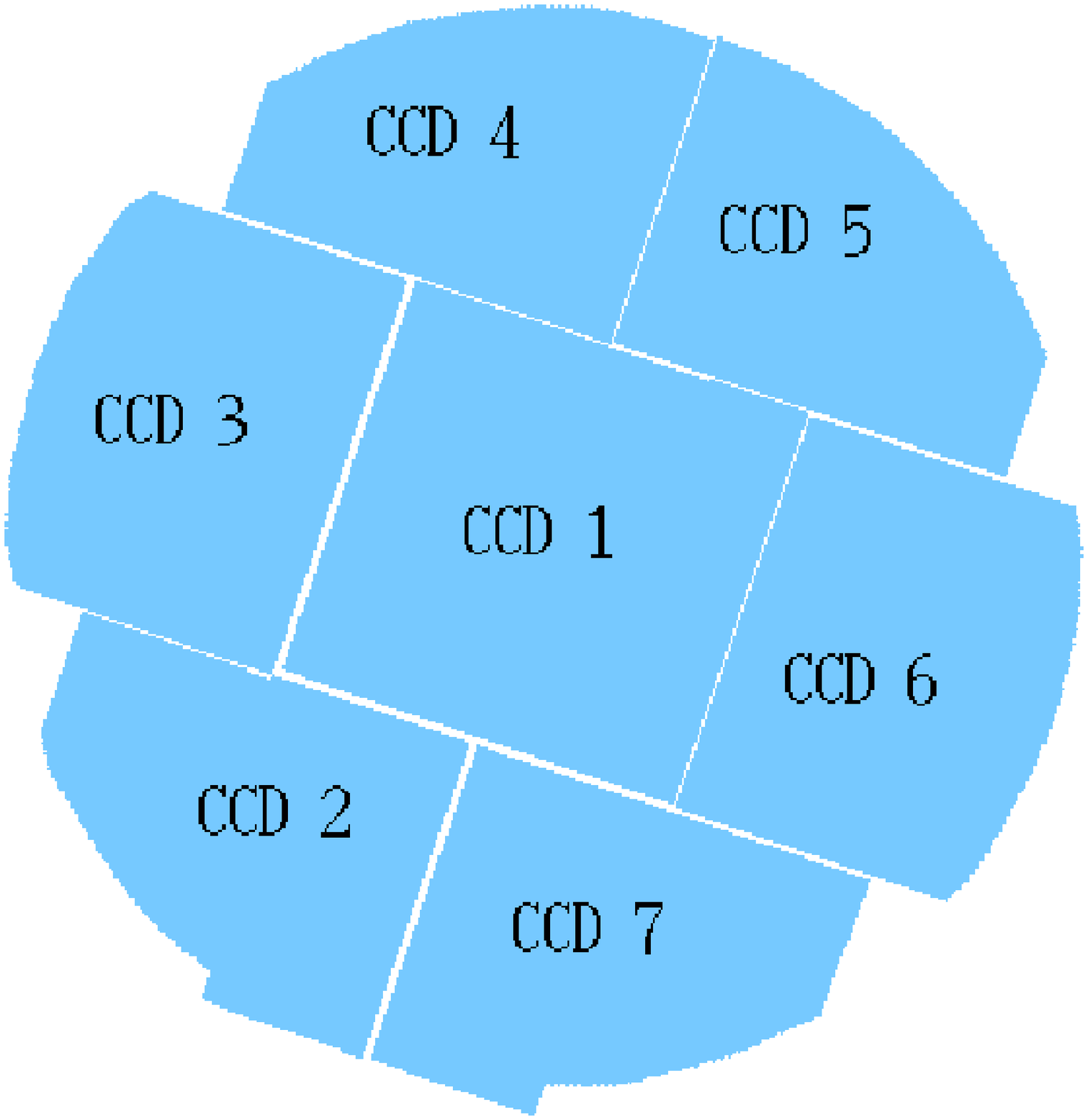}
\caption{Position of the CCDs for each EMOS camera (EMOS 1 to the left, EMOS 2 to the right) with the orientation of the HCG1 sky coordinates; north is up the east is to the left}
\label{fig:ccdposition} 
\end{figure}

\begin{itemize}
\item The two noisy CCDs of EMOS 2 (CCD2 and CCD5) have not been excluded during the wavelet analysis in order to have a more continuous image. On the one hand, as our electronic noise cleaning is not perfect, the final result is an artificially high signal in the corresponding regions (see Fig. \ref{fig:ccdposition}). This explains the high flux to the north-west (CCD5; note how this higher emission follows the shape of the CCD) and to the south-east (CCD2). On the other hand,  for one exposure, CCD2 was turned off, and thus the total (EMOS 1+EMOS 2) image has a lower S/N ratio in the south-east zone corresponding to the EMOS 2 CCD2. The net effect is that, in this region, the structure detection will be less efficient (our wavelet algorithm does not allow for the moment to give a different weight to different spatial regions), and thus the smoothing kernel will appear larger (in this region) in the reconstructed wavelet image. Discarding the noisy CCDs would noticeably increase this effect.

\item The  high signal to the  west-north-west is further increased by a point source highly degraded by the off-axis PSF;

\item The extension onto the south-east (in the direction of galaxy d) is also 
probably due to a radio source (DSM99). This source is detected on CCD2 of EMOS 2, but because of the high CCD temperature in the early stage of the calibration, there is a great deal of electronic noise which effectively prevents further investigation of the source.

\end{itemize}

\subsection{The radial profile}

The gas density distribution, obtained from the surface brightness
profile, is one of the parameters (together with the temperature and
total flux) used to measure the gravitational potential well in the
assumption of spherical symmetry.

We define a mask excluding all sources detected in the wavelet image as well as detector artifacts such as bad lines/columns and hot CCDs not taken into account by the pre-processing. Because of the uncertainties in the
spatial distribution at the energies corresponding to the Al and Si
fluorescence lines in the particle image (see section 2), we again
limit our analysis to the 0.2-1.4 keV energy
band. We are confident that this choice does not change our result
because the preliminary spectral analysis (section 3.1) showed that
the X-ray diffuse emission has its maximum around 1 keV. 

\begin{figure}[h]
\begin{center}
\includegraphics[width=8.cm,height=5cm]{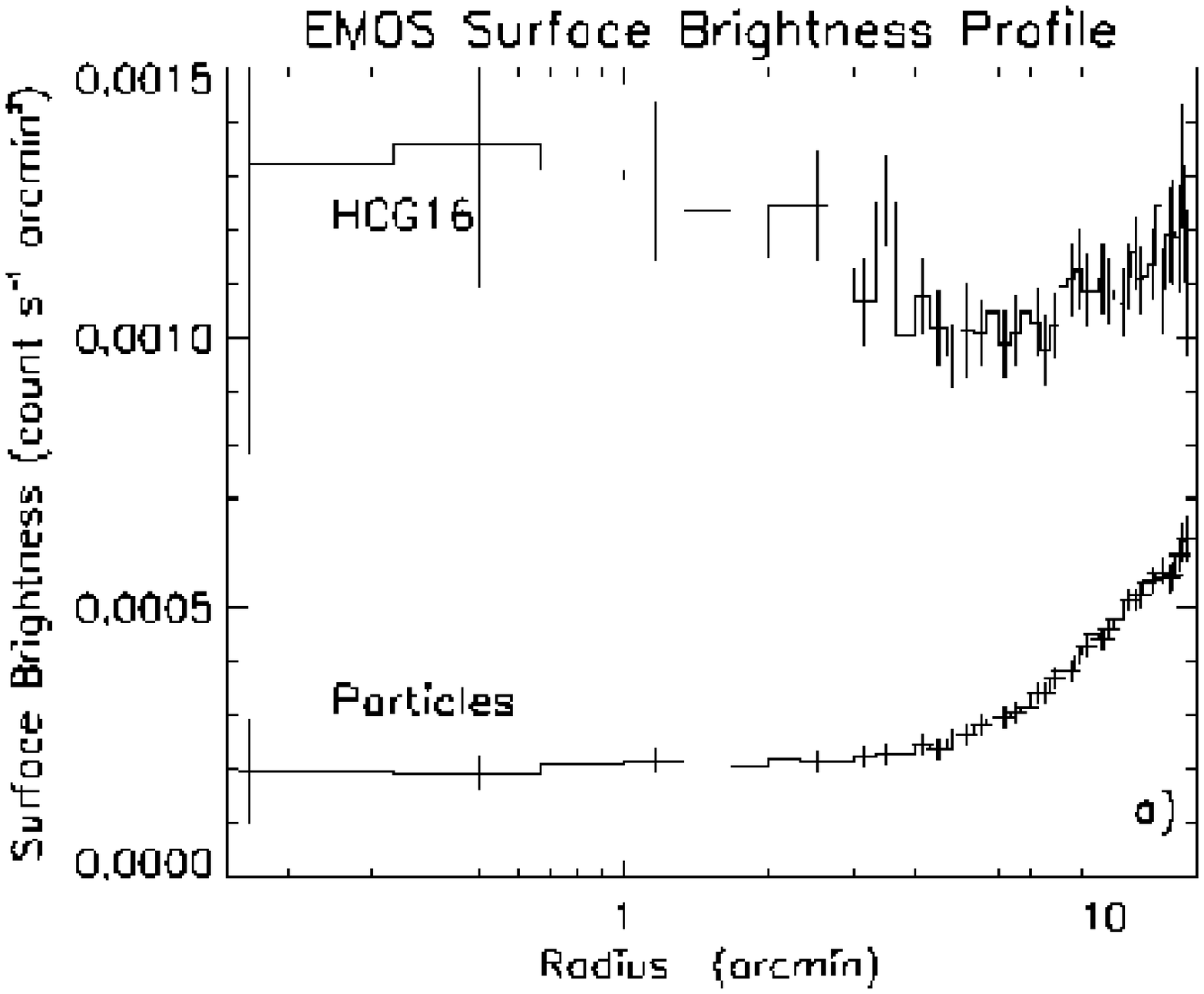}
\includegraphics[width=8.cm,height=5cm]{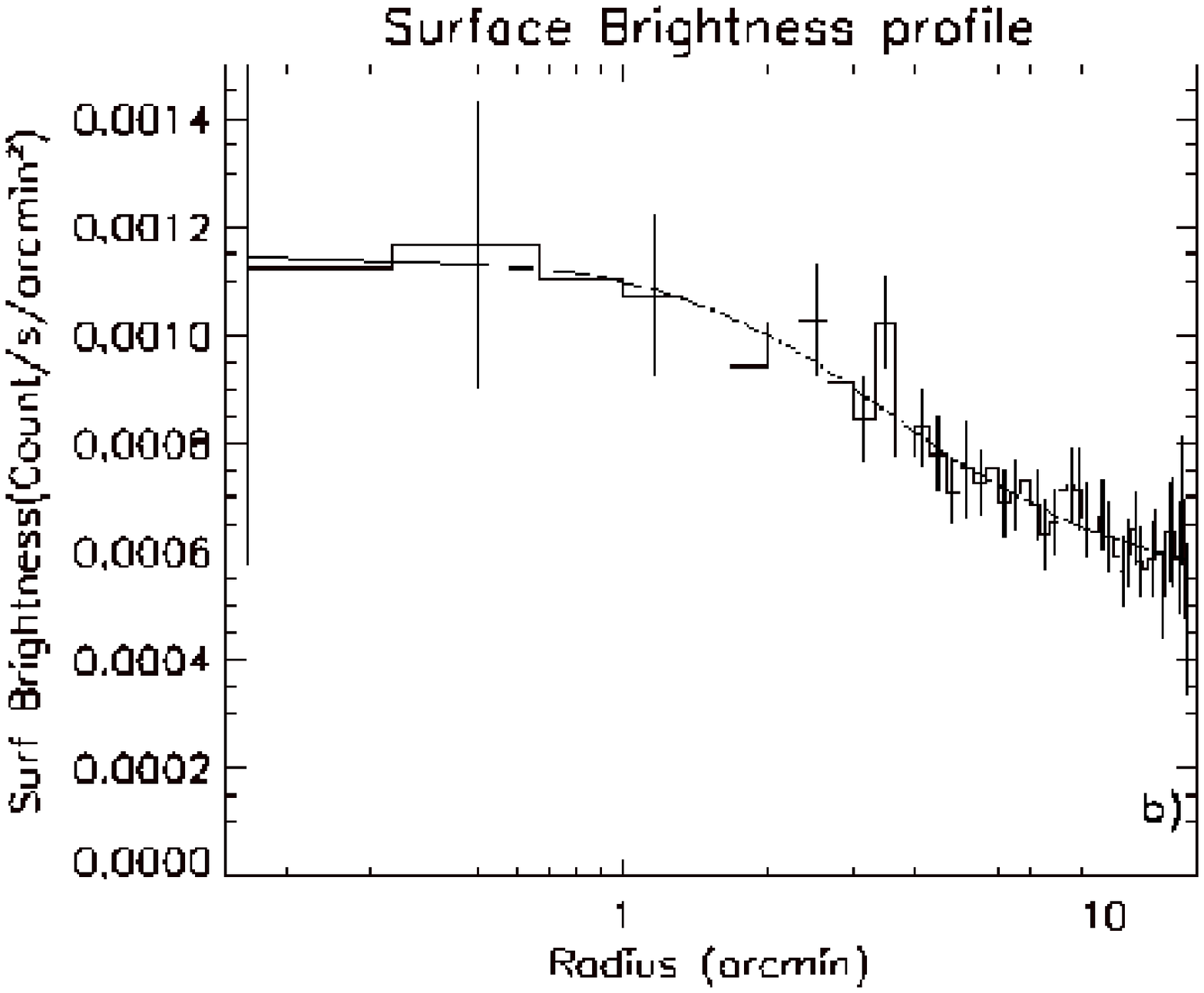}
\caption{{\bf a)} The surface brightness profile for the HCG 16 (EMOS 1+EMOS 2) and for the simulated particle image in the 0.2-1.4 keV energy band. Both profiles have been calculated from images corrected for vignetting. The two radial profiles are computed using the same mask and in the same energy range. {\bf b)} The EMOS 1+EMOS 2 surface brightness  profile after subtraction of the particle background. The parameters of the $\beta$ model fit are listed in table 2.}
\label{fig:radprof}
\end{center}
\end{figure}

To correct for vignetting, we used the photon weighting method described in Arnaud et al. (2001): each photon is weighted by the ratio of the effective area at its position on the detector to the central effective area at the energy of the photon. To take into account the ``false'' vignetting correction of the particle background component, we also corrected the particle image for vignetting, in the same energy band, and with the same algorithm. In the energy band considered here, the contribution of the particle background is, in principle, negligible. However, because of the telescope vignetting, in the outer regions more than half of the X-ray photons are lost, while the particle flux stays constant across the whole FoV. Thus the  particle background contribution must be subtracted because, as shown in Fig.~\ref{fig:radprof}, at large radius ($>$ 7\arcmin) this component contributes $\sim 50\%$ of the total sky background.

\noindent The particle image was normalised, as described in section  2, to  the 10 to 12 keV energy band count rate. This was done separately for EMOS 1 and EMOS 2 and only after normalisation  were the two images added together.

\noindent The surface brightness profile was obtained  for the normalised (EMOS 1+EMOS 2) particle image and for the HCG 16 image separately (Fig. \ref{fig:radprof}a), but using the same mask to exclude point sources. Afterwards, the particle radial profile was subtracted from that of  HCG 16  (Fig. \ref{fig:radprof}b). To summarise, if $I = E \times S + P$ where $I$, $S$, $P$ and $E$ indicate the observed image, the source image, the particle image and the exposure map (including vignetting), respectively, then the whole radial profile analysis can be written as $S = I/E - P/E$.

An estimate of the mass of the diffuse intergalactic medium is given
by fitting the radial profile with a King profile ($\beta$
model). However the HCG 16 gas shape (Fig. \ref{fig:fig2}) indicates that the
$\beta$-model method assumption of spherical symmetry is not
necessarily true. The centre of the fit was obtained, after several iterations, in order to maximise the central X-ray surface brightness of the King profile. The best solution was $\alpha = 02^h09^m32.0^s; ~\delta= -10^o09\arcmin00\arcsec$ (J2000), which corresponds within 0.3\arcsec~ to the optical centre (given by SIMBAD, FK5 coordinates). The best $\beta$-model fit of the surface brightness profile gives a $\beta$ parameter of 0.37, a core radius of 2\farcm.7 and the central intensity of 6.5$\times$10$^{-4}$ cts s$^{-1}$ arcmin$^{-2}$
(Table 2).  To check the consistency of this result and to have an
estimate of the statistical significance of the error bars (given at 1
$\sigma$ confidence level), we determine the background level
independently, i.e. computing the count rate in the [0.2 -1.4]  keV energy
band and in the 10\arcmin~ to 12\arcmin~ ring (we consider it safer to limit
the large radius analysis to 12\arcmin~ since outside this radius some
systematic error may come from the vignetting function, which is poorly
known beyond this radius). This gives a value of $6.0 \times 10^{-4}$ cts
s$^{-1}$ arcmin$^{-2}$ which is consistent with the value found when
we fit the profile with the background parameter left free. However,
due to the very low S/N, the fit is very sensitive to the background
level. This may have an important effect on the fitted parameters,
and also in the mass determination. 
The poor constraints on
the fitted parameters are thus not surprising and we have to be
careful with regard to their use in the determination of physical
properties.


\begin{table}[ht]
\caption{The $\beta$ model fit results. Errors are quoted at 1 $\sigma$ significance}
\begin{tabular}{lc}
\hline
Parameter & fit value \\
\hline
$\beta$ & $ 0.37 \pm 0.3$ \\
$R_c$ (arcmin) & $2.7 \pm 2.5$\\
I$_0$ (cts s$^{-1}$ arcmin$^{-2}$) & ($6.5 \pm 0.9$) 10$^{-4}$ \\
bkg (cts s$^{-1}$ arcmin$^{-2}$) &  ($5.0 \pm 3.0$) 10$^{-4} $ \\
\hline
\end{tabular}
\end{table}

\subsection{The spectra}
\subsubsection{Method}
While a standard technique for extracting spectra is to consider a
circle around the source centre, we use a different method to
determine the spatial extraction region due to the shape of the
detected signal (Fig. \ref{fig:fig2}), the low surface brightness (Fig. \ref{fig:radprof}b) and
the low S/N of the observation.
Assuming that the wavelet analysis is the best representation for the
detection of the diffuse emission, we define a regular elliptical region, following the shape of the emission in the wavelet image (Fig. \ref{fig:fig2}).
Within this elliptical mask, all bright sources have been excluded by defining a threshold (1 cts/s) in
the wavelet image, which results in a typical radius of  1\farcm.1 for the exclusion of the galaxies. We will refer to this elliptical mask region as region $E$. 

\noindent Given that the intragroup gas is detected mainly in the central CCD (Fig. \ref{fig:fig2}), it is possible to estimate a local background in an external region corresponding roughly to a 10\arcmin~ to 12\arcmin ~ring around the centre of the FoV (Fig. \ref{fig:fig2}). We will refer to this local background as region $R$.

 To avoid the introduction of systematic errors, we are, in this spectral analysis, more conservative than in the imaging analysis and in both regions $E$ and $R$ we exclude: 1) the south-east corner because of contamination due to CCD2 of EMOS 2, which has a much higher noise (see section 2); 2) the region corresponding to CCD5 of EMOS 2, the other CCD displaying higher noise and which shows excess emission at low energies; 3) in both cameras, the region to the WNW because of a highly PSF degraded point source (note that this region corresponds essentially to CCD4 in EMOS 1). 

The false vignetting correction of the particle component of the local background will over-subtract the true background contribution. To avoid this error, the events are not initially corrected for the vignetting effect. The method used follows these steps:

\begin{enumerate}
\item We extracted 2 spectra from each events list (HCG 16 and particle): the first in region $E$, the second in region $R$. We now have spectra $E_s, E_p, R_s, R_p$ where the $s$ spectra are extracted from the HCG16 event list and the $p$ spectra come from the particle event list. 

\item The particle spectra were subtracted from the HCG 16 spectra in each region after a normalisation ($n$) in the 10 to 12 keV band:

\[
E_x = E_s -nE_p
\]
\[
R_x = R_s -nR_p
\]

This gives the particle subtracted spectra $E_x$ and $R_x$.

\item {\it Vignetting correction}: in order to compute the vignetting correction factor (VCF), we work under the assumption that the spatial distribution over the whole integrated region is independent of the energy (i.e. the spectrum is similar throughout the region). We thus use all photons, irrespective of their energy, to
describe the spatial distribution of events. Then, for each photon
energy, we compute the vignetting factor by averaging the weight (see
Sect. 2 for the weight definition) at each photon position. The
advantage of this method is that in using the total number of photons
to calculate the weight, we can improve considerably the accuracy of
the calculated vignetting factor.

\noindent Spectra $E_x$ and $R_x$ were thus multiplied by their respective vignetting correction factors (VCFs):

\[
E_{x,vcf} = E_x \times {\rm vcf_E}
\]
\[
R_{x,vcf} = R_x \times {\rm vcf_R}
\]

\item The spectrum $R_{x,vcf}$ was then normalised to the surface area of region $E$, and subtracted:

\[
E_{\rm net} = E_{\rm x,vcf} - mR_{\rm x,vcf}
\]
where $m$ is the normalization. 
This gives the full background-subtracted, vignetting-corrected net spectrum, $E_{\rm net}$.
\end{enumerate}
 
\subsubsection{PSF considerations}
An important point of concern is the contamination of the excised sources due to the wings of the EMOS PSF.
The one dimensional EMOS PSF is well described by a King function out to 5\arcmin ~radius, and in the energy range of interest here, is essentially energy-independent (Ghizzardi 2001). The asymptotic slope of the PSF model is $\alpha= 1.46$ and $\alpha=1.41$ for EMOS 1 and EMOS 2 respectively, at 0.8 keV in the on-axis position. We adopt this description of the PSF to correct the spectrum of the diffuse gas for the PSF of the excised sources (galaxies and point sources) in the ellipse and background (ring) regions.

In determining the spectrum in the elliptical region, we have excised the sources using a typical radius of $1\farcm.1$ for the galaxies and $24\arcsec$ for the point sources. The sources in the background ring were excised in a cutoff radius of typically 40\arcsec.

For each source we calculated the encircled energy fraction (EEF) at 0.8 keV (corresponding to the peak of the diffuse emission spectrum) in the circle used to excise it, taking into account its off-axis position. The fraction of this energy falling outside the excised circle is $F= (1-EEF)/EEF$. Once the fraction for each source was obtained, we computed a statistical mean fraction $\langle F \rangle$ for the two regions of interest, the ellipse and the background. The mean fraction $\langle F \rangle$ is 0.27 for EMOS 1 point sources in the ellipse, for example. This means that 0.27 of the flux of the excised point sources contaminates the diffuse X-ray gas flux of EMOS 1. 

The PSF correction for the galaxies is more subtle because, in principle, we have to take into account the extension of a galaxy and convolve the emission distribution with the PSF. We thus use the EMOS PSF model to investigate the systematic errors introduced by treating a galaxy  as a point source and  as modelled by a $\beta$ model. We applied this test to HCG16c, the brightest and most extended galaxy, and thus the most extreme case. We obtained a surface brightness (SB) profile of this galaxy from the EMOS 1 image in the energy range 0.2-1.4 keV and we fitted it with the model of a point source convolved with the EMOS PSF. We fit this profile up to 2\arcmin ~after which the profile becomes dominated by the diffuse emission. The fit is not good, but if we extrapolate the model up to 5\arcmin~ we obtain a $F$ = 0.063 using a cutoff radius of 1\farcm.2 for this galaxy. 

\noindent We then obtained a  SB profile of galaxy HCG16c from the Chandra image in the same energy range as above. The Chandra PSF can be considered as perfect with respect to the EMOS PSF; we thus assume that this profile represents the intrinsic profile of the galaxy. We fitted the Chandra SB profile with a $\beta$ model finding as a best fit $\beta = 0.75^{+0.08}_{-0.06}$, $R_c = 5\farcs28^{+1\farcs32}_{-1\farcs02}$ and a reduced $\chi^2$ of 25.37/25 $\simeq 1$.

To estimate the EMOS spread function we fitted the EMOS profile (up to 2\arcmin) with  a $\beta$ model convolved with the EMOS PSF and with the $\beta$ and the R$_c$ fixed to their respective Chandra best fit values. The central X-ray surface brightness and background were free parameters. In this case we obtain $F$ = 0.07. If we leave the $\beta$ parameter free, we obtain  $\beta = 0.81\pm0.04$ for a reduced  $\chi^2$ of 1.18. This value is in good agreement with the value obtained from the Chandra profile. In this case, the obtained $F$ is 0.068. 

To summarise, if we consider this galaxy as a point source, we make a maximum systematic error of 10\%. This induces an error on the X-ray luminosity obtained from the spectrum of the diffuse gas of $\sim$ 1.5\%, which is consistent with the statistical errors.
In view of the results outlined above it appears to be justified to treat galaxies as if they were point sources. We computed a mean $\langle F \rangle$ value for all galaxies (per camera) obtaining $\langle F \rangle$ = 0.073 for EMOS 1  and  $\langle F \rangle$ = 0.089 for EMOS 2. The mean cutoff radius was 68\arcsec~ for both cameras. In the calculation of the EEF the off-axis position was taken into account for each galaxy; we note however that this does not introduce a strong effect.

We extracted the global spectrum  (the best representation of the total flux) of all point sources in the background ($PR$; ring) region and in the ellipse ($PE$) region. The global spectrum of the 4 galaxies was also obtained ($G$). We then calculated a background spectrum $B_c$ which takes into account not only the PSF contamination due to its own point sources, but also that due to the point and extended sources in the elliptical region. In other words:
\begin{equation}
B_c = B - \left(PR \times \langle F\rangle_{PR}\right) + \left(PE\times \langle F \rangle_{PE} + G\times \langle F \rangle_G\right)
\label{eq:bkgpsf}
\end{equation}

\subsubsection{Spectral fit results}
The EMOS 1 and EMOS 2 spectra were binned to 2$\sigma$ statistical significance after background subtraction, and fitted with an absorbed MEKAL model (XSPEC v11.0), the column density being fixed to the galactic value ($2.0 \times 10^{20}$ cm$^{-2}$), leaving the temperature, abundance and normalisation (emission measure) as free parameters. 

\begin{figure}[h]
\begin{center}
\includegraphics[width=5.5cm,height=8.5cm,angle=-90]{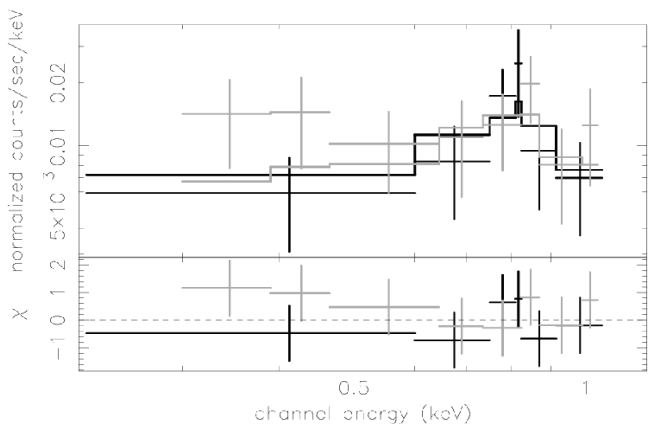}
\caption{a): EMOS 1 (black) and EMOS 2 (gray) background subtracted PSF corrected spectrum, folded with the combined fitted  model. The background subtracted spectra are binned to the  2$\sigma$ level. The channels corresponding to  0.2 to 0.3 keV were  ignored for EMOS 2 because of low energy noise not taken into account in the data processing (cf. CCD temperature).}
\label{fig:fig4}
\end{center}
\end{figure}

\noindent The EMOS 1 and EMOS 2 results are in very good agreement, except in the 0.2-0.3 keV energy band, where the EMOS 2 shows  excess emission. We attribute this difference to an instrumental effect not well taken into account in the data processing (probably due to the contamination of the central CCD by the high temperature CCDs nearby, particularly affecting the low energy band) and decide consequently to ignore, when fitting, the 0.2-0.3 keV band  in EMOS 2. We used the response matrix version 20.6. In Fig. \ref{fig:fig4} the EMOS 1
and EMOS 2 spectra and the folded model are shown. In the energy range 0.2-1.4 keV the EMOS 1 camera provides 1635 counts, corresponding to 313 net counts after background subtraction, while the EMOS 2 measured counts are 1269 in the energy range 0.3-1.4 keV (327 net counts).
 The spectral fit results are listed in Table 3, which also lists the fit results before the correction for the PSF. We can notice that the correction for the wings of the PSF, obtained by using $B_c$ instead of $B$ for the background spectrum (see eq.\ref{eq:bkgpsf}), has the effect of slightly reducing the temperature, but above all of reducing the flux by $\sim$20\%. The relative contribution of each galaxy (treated as a point source) and total point source contribution to this 20\% flux reduction is shown in detail in Table \ref{tab:percentcont}.

\begin{table*}[ht]
\caption{Spectral fit results before and after PSF correction. {\em Top}:  Column (1): absorbing hydrogen column density, we adopt the Galactic value from Stark et al. (1992). Column (2): temperature. Column (3): metallicity. Column (4): $\chi^2$ of the spectral fit. Column (5): redshift. {\em Bottom}: Column (1): X-ray luminosity computed in the PSPC energy band. Column (2): bolometric X-ray luminosity, uncorrect for absorption. Column (3): bolometric X-ray luminosity, uncorrect for absorption and corrected $^*$ for the omitted emission from the galaxy regions (follow the discussion in Sect. 4.1). Column (4): bolometric X-ray luminosity corrected for absorption. Column (5): bolometric X-ray luminosity corrected for absorption and corrected as in column (7)$^*$. Errors are at the 90\% confidence level. The error on the luminosity has been obtained by fixing the metallicity parameter range in the fit to be between 0 and 2 solar.}

\begin{center}
\begin{tabular}{|ccccc|}
\hline
      &                      &  no PSF corrected  &  & \\
\hline
\hline
$N_H$ &                   $kT$ &   $Z/Z_{\odot}$ & $\chi^2$/d.o.f.& $z$ \\
(10$^{22}$ cm$^{-3})$  & (keV)  &              &                &   \\
\hline
0.02 (fix) 	       & 0.53 $_{-0.13}^{+0.11}$ & 0.09 $_{-0.04}^{+0.12}$ & 13.84/20 & 0.0132 \\
\hline
      &                      &   $L_X$ (10$^{41}$ erg s$^{-1}$) &  & \\
0.2-2.3 keV &  bol. uncorr.  & bol. uncorr. (*)                 & bol. corr. & bol. corr. (*)\\
\hline
0.51 $_{-0.07}^{+0.05}$ &0.53$_{-0.07}^{+0.07}$ & 0.69 & 0.92 $_{-0.11}^{+0.13}$ & 1.2 \\ 
\hline
\hline
      &                      &   PSF corrected &  & \\
\hline
\hline
$N_H$ &                   $kT$ &   $Z/Z_{\odot}$ & $\chi^2$/d.o.f.& $z$ \\
(10$^{22}$ cm$^{-3})$  & (keV)  &              &                &   \\
\hline
0.02 (fix) 	       & 0.49 $_{-0.16}^{+0.19}$ & 0.07 $_{-0.05}^{+0.31}$ & 6.35/11 & 0.0132 \\
\hline
      &                      &   $L_X$ (10$^{41}$ erg s$^{-1}$) &  & \\
0.2-2.3 keV &  bol. uncorr.  & bol. uncorr. (*)                 & bol. corr. & bol. corr. (*)\\
\hline
0.39 $_{-0.08}^{+0.08}$ &0.40$_{-0.08}^{+0.08}$ & 0.52 & 0.72 $_{-0.12}^{+0.12}$ & 0.96 \\ 
\hline

\end{tabular}
\end{center}
\end{table*}

\begin{table}[ht]
\caption{Relative contribution of each galaxy and total point sources to the 22\% flux reduction due to the PSF contamination. The contribution of each galaxy has been obtained by integrating the total counts in the cutoff radius (in the energy range 0.2 -1.4 keV for EMOS 1 and 0.3-1.4 keV for EMOS 2) used to excise it and by multiplying this value for the mean PSF correction factor $\langle F \rangle_G$. HCG16a and HCG16b are considered together. The total point source contribution has been estimated  using $\langle F \rangle_{PE}$ as the PSF correction factor.}
\begin{center}
\begin{tabular}{ccc}
\hline
\hline
region & EMOS 1 & EMOS 2 \\
	& \% & \%        \\
\hline
HCG16a+ HCG16b& 23.6 & 22.7\\
HCG16c & 23.8 & 25.5\\
HCG16d & 7 & 8.8\\
point sources (total) & 45.5 &  43\\
\hline
\end{tabular}
\end{center}
\label{tab:percentcont}
\end{table}

\section{Discussion}
\subsection{Comparison with previous works}
The image analysis presented in Sect. 3 and the surface brightness
 profile confirms that HCG 16 is a bound system as previously shown from ROSAT studies. The EMOS sensitivity is 1.8 times that of the PSPC instrument on-board ROSAT at low energies, and the useful XMM-Newton observation time is more than triple that of the ROSAT/PSPC observation. This allows a better source detection (while point sources are in the noise in the PSPC observation) and exclusion of the galaxy contribution and the establishment of
 stronger constraints on the temperature.  On the other hand, the high
 background level of XMM (worsened by the electronic noise at the
 early stage of the instrument setting) reduces the S/N and the
 detection radius for extended sources. Relative to ROSAT/PSPC, the EMOS PSF wings are larger, leading to a smoother apparent emission than was found by DSM99. In our analysis we have taken into account all these major sources of contamination, finding that the results are stable in the limit of this first light observation. The diffuse emission extends
 to at least 6\arcmin ~from the group optical centre (corresponding to
 $\sim$ 135 kpc at the distance of HCG 16). Our results show a smooth
 gas distribution, rather in contrast to  the very irregular morphology
 noted by DSM99. These authors emphasise the nature of region C4, corresponding to an excess emission starting from galaxy HCG 16b toward the south.
A quick comparison between the ROSAT/PSPC and the EMOS image shows that the emission to the  south-western extreme of our ellipse corresponds to a point source which was marginally detected by ROSAT and which is completely resolved by EMOS. If we smooth the EMOS image in a similar way as done by DSM99, we do not find a spatially continuous signal between the point source (widened by the large Gaussian kernel) and the diffuse emission. We thus estimate that at least 1/3 of the C4 emission is due to a point source unrelated to the group. While the statistics are poor, the spectrum of this source is minimised by a power law of $\alpha = 2.25$ and a column density of $6.7 \times 10^{19}$ cm$^{-2}$. The flux of the source is 1.7$\times10^{-14}$ ergs cm$^{-2}$ s$^{-1}$ and if the source was at the HCG 16 redshift its total luminosity would be 1.3$\times10^{40}$ erg s$^{-1}$. The remaining 2/3 of the flux are effectively inside the ellipse in Fig. \ref{fig:fig2}, and there is no clear sign of separation from the neighbouring regions of the diffuse emission. An additional source lies in the region corresponding to C4 in DSM99 (closer to galaxy b). It has been excised in our analysis, but we checked the contribution of this source to the C4 emission. The best fit is a power law of index 2.3, with a flux of $5.3\times 10^{-15}$ erg cm$^{-2}$ s$^{-1}$.

 The elliptical shape and the spatial
 extension of the entire gas distribution make it unlikely that this gas can
 come only from ejection by the galaxies: even if active galaxies (as
 is the case here) eject large amounts of gas via galactic winds
 (Fabbiano et al. 1988), this emission does not extend out to several
 galactic radii. 

 While HCG 16 still remains one of the coolest galaxy groups with 
detected diffuse X-ray emission, the EPIC/MOS spectra analysis of
HCG 16 gives a temperature of 0.47 keV, higher than that obtained with
ROSAT (PBEB, DSM99), but in agreement within statistical errors.

\noindent The best fit metallicity is 0.07 solar, which is low with
respect to galaxy clusters and some galaxy groups. However, even 
though several authors have reported lower abundances in groups than in
clusters (e.g. Davis et al. 1999), the reality of
this result is not well established, and
we have to be careful with abundance determinations of such a low
temperature systems. The abundance derived from hot plasma
model fits for systems with
temperatures below $\sim$1 keV is primarily driven by the iron L-shell
complex, and the combination of the different lines is very sensitive
to the temperature. This means that if temperature gradients are
present, the abundance is, in general, an underestimate (Buote \&
Fabian 1998; Finoguenov \& Ponman 1999). Works based on ROSAT/PSPC
(Mulchaey \& Zabludoff 1998) and ASCA (Finoguenov \& Ponman 1999) data have 
already pointed out the multi-temperature structure of several galaxy
groups. We are unable to determine a temperature profile for HCG 16
with these data, but given the presence of several starburst galaxies interacting in the central region it seems unlikely that the gas will be isothermal.  On the other hand, our best fit metallicity
determination is poorly constrained. Moreover the
metallicity depends on the model used and it has also been shown, for
other groups, that an increase of the metallicity value is obtained if
a 2 temperature model is used (Buote 2000). In any case, we find a non-zero metallicity.

The low value of the bolometric X-ray luminosity is not
surprising. Galaxy groups, especially at low temperature, have flat
surface brightness distributions (Ponman et al. 1999; Helsdon \&
Ponman 2000a). In this work, the X-ray luminosity is calculated from
the emission measure of the MEKAL model used to fit the spectrum. The
latter is extracted within the radius where significant X-ray emission
is detected, thus the luminosity calculated here may be only a small
fraction of the total luminosity.

\noindent For comparison, we consider the whole region C2+C3+C4 in DSM99, taking the metallicity as 0.1 solar. Our result 
(0.52$^{+0.1}_{-0.1}$ 10$^{41}$ \es) and theirs (0.69 10$^{41}$ \es) do not strongly disagree if we consider that $\sim$ 1/3 of  of the X-ray emission form C4 comes from a point source, and that the spatial regions and the energy band where the spectrum is extracted are not
exactly the same. 

In comparison to the PBEB result, $L_X= (4.7\pm 1.1)~10^{41}$ \es, the
bolometric X-ray luminosity found in the present study is more than a factor 5
 lower (4.7/0.72). There
are several reasons for this. According to DSM99, a factor of 2 of the luminosity found by PBEB is attributable to foreground/background sources, while the good EPIC sensitivity allow us to excise some sources missed in the PSPC analysis by these authors. Our limited ability to detect diffuse emission at large radii due to the high EPIC background is another reason for the difference in the surface used to extract the spectrum, although DSM99 argue that, apart from region C4 there is no emission far from the galaxies. Moreover, in our work we  use a different correction for absorption relative to PBEB because of our higher temperature. 

\noindent  Furthermore, we have to consider that in addition to the fact that PBEB used a larger  region to extract the spectrum (200 kpc), the flux from the spatial region occupied by the galaxies has been omitted in our analysis, whilst PBEB included an
estimate of the diffuse group flux underlying the galaxy
components. Hence the total luminosity within the elliptical region
considered represents a lower limit to the total diffuse flux of
the system.  Assuming that the area omitted behind the galaxies
has the same mean brightness as the rest of the region, we calculate
the ratio between the mask area and the whole elliptical area (Fig.
\ref{fig:fig2}). This factor gives a rough correction of 30\% to be applied to the
$L_X$ value leading to an estimate of 9.6 10$^{40}$ \es ~within the
full elliptical region. To investigate further the comparison with
ROSAT results, we analysed the PSPC data within a 5\arcmin~ radius of
the EPIC detection. Computing the count rate in this circle (which is
comparable to our mask region) and replacing the galaxy emission with
the interpolated flux (as done in PBEB), we obtain a bolometric
luminosity of 7 10$^{40}$ \es, thus the results from EPIC and
ROSAT, within similar central areas, are in an acceptable agreement.

\subsection{Mass estimates and scaling laws}
\begin{figure}[h]
\begin{center}
\includegraphics[width=8.5cm,height=6.5cm]{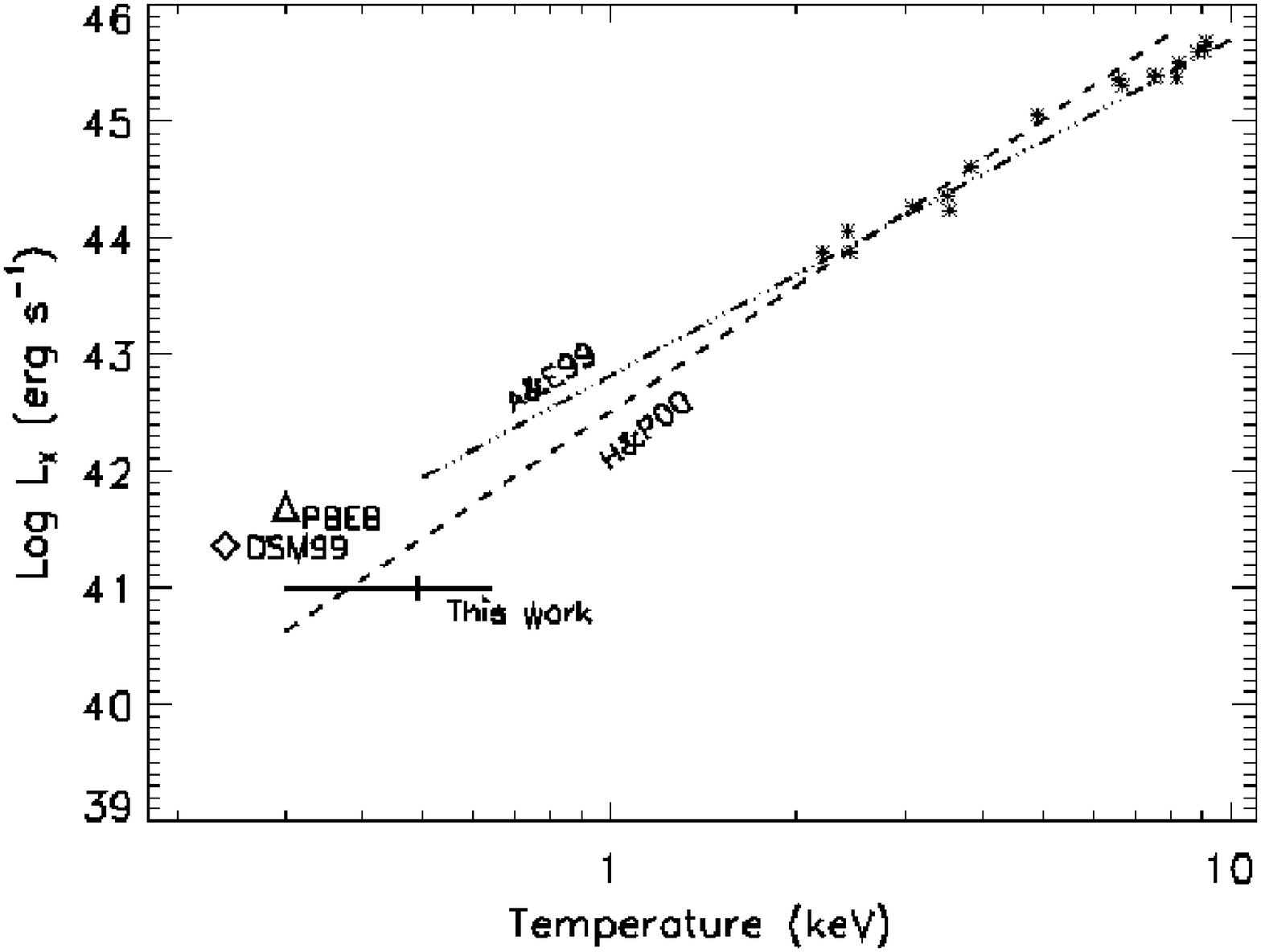}
\caption{$L_X$ - T relation for cluster of galaxies (from Arnaud \& Evrard 1999 - H\&E99) compared to the one for group found by Helsdon \& Ponman 2000 (H\&P00). Our result is marked as well as the DSM99 (diamond) and PBEB (triangle) results.}
\label{fig:fig5}
\end{center}
\end{figure}

 An estimate of the total mass of the gas can be obtained by combining the $\beta$-model parameters and the spectral results. This implies that we assume, despite our reservations (section 3.4, 3.3)  that the gas is isothermal and has a spherical symmetry. Under these hypotheses, the mass of the gas amounts to  8.8$\times$10$^{10} M_{\odot}$ at the maximum radius of detection (135 kpc). 
If the mass profile is extrapolated out to the virial radius ($\simeq$ 900~kpc)\footnote{the virial radius is obtained as: r$_{vir}$ = 4.1 (T/10
kev)$^{0.5}$ ~ (1+z)$^{-1.5}$ ~ $h^{-1}_{50}$ (Evrard et al. 1996)}
the total gas mass is 3.7$\times$10$^{12} M_{\odot}$. We stress that we have  to be careful about this result because we have made several
assumptions which may not be strictly valid. Simulations indicate that
the $\beta$ value obtained fitting the surface brightness profile is
very sensitive to the range of radii used in the fit (Navarro et
al. 1995; Vikhlinin, Forman \& Jones 1999) and in the case of very low surface brightness the
background determination is a strong source of uncertainty.

Self-similar scaling laws valid for clusters of galaxies break down at
the low mass limit (PBEB; Helsdon \& Ponman 2000a, Xue
\& Wu 2000). In the isentropic limit, semi-analytical models (Cavaliere
et al. 1997, Balogh et al. 1999) predict a change in the slope of
the $L_X$-T relation at a temperature around 1 keV, which is in
agreement with the observational results from galaxy groups (PBEB,
Heldson \& Ponman 2000a, 2000b; Mulchaey 2000). This departure from
self-similarity of low temperature systems can be explained by
preheating models, in which the gas is heated by some non-gravitational
effect (such as galaxy winds during the galaxy formation) before it
falls into the potential well of the group (Helsdon \& Ponman 2000a,
Babul et al. 2002, Tozzi et al. 2000; Dos Santos \& Dor\'e 2002).  In this context, HCG 16 becomes
even more interesting: its low surface brightness, low temperature
and low velocity dispersion locate this group in an extreme
position in all scaling laws determined for clusters and
groups. Figure \ref{fig:fig5} shows the $L_X$-T relation from Arnaud \& Evrard
(1999) for clusters, together with our result, where we take the bolometric luminosity corrected for absorption and for the omitted region corresponding
 to the galaxies. The dot-dot-dashed line is
the best fit for clusters, whereas the dashed curve is the best fit
obtained by Helsdon \& Ponman (2000b) using the same sample as PBEB (22
groups observed with ROSAT/PSPC) but considering a different
statistical weight when fitting the $L_X$-T relation. HCG 16 is a
factor 10 lower with respect to the cluster relation. However, despite
its extreme nature, it is in reasonable agreement with the correlation found
for galaxy groups,  especially if one considers  that integration
to a larger radius  
could increase $L_X$ by a factor of a few.
 One should also bear in mind that the $L_X$-T relation 
for groups shows
significant scatter (Helsdon \& Ponman 2000b), interpreted as arising
from the different formation history of individual groups.

Another manifestation of the breaking of self-similarity seen
in groups is the progressive flattening in X-ray surface
brightness profiles in cooler systems (Ponman et al. 1999).
For a system with a mean temperature as low as that of HCG 16,
this trend implies a surface brightness index $\beta$ less than 0.4
(Helsdon \& Ponman 2000a). Our fitted value, $\beta=0.37 \pm 0.3 $, 
(Table 2) is therefore well-matched to the trend seen in other
X-ray bright groups.

Combining our fitted gas density profile with the measured gas temperature
(which we take to be isothermal), it is possible to derive an entropy
profile for the intergalactic gas. The result is a rising profile, with
a value, at $r<6\arcmin$ of $\sim$100~keV~cm$^2$. Although the profile itself
should not be taken too seriously, due to uncertainties in the 
radial trends arising
from the limited spatial extent over which we can reliably trace
the emission, the entropy derived in the inner regions should be
a secure estimate. This is very similar
to the `entropy floor' value found in the inner regions of X-ray bright 
groups by Ponman et al. (1999) and Lloyd-Davies et al. (2000).

\subsection{The nature of HCG 16}
Although the quality of this first light EPIC observation is not all
we might wish, and we are therefore unable to trace the diffuse
emission from HCG 16 out to even as large a radius as was reached
with ROSAT, the superior  spectral resolution and sensitivity 
of XMM, compared to the ROSAT PSPC, allow us to establish
some important facts about the group:
\begin{itemize}
\item There is substantial, genuinely diffuse X-ray emission
in the centre of the group, surrounding the galaxies and extending
to a radius of at least 135 kpc from the centre of the group.
\item The mean spectrum of this emission is well represented
by a thermal plasma with a low temperature (0.49~keV) and non-zero
metallicity.
\item The total extent and luminosity of this hot gas suggests
that it cannot be attributed to galactic wind emission
from the active galaxies within the group, and the morphology
of the gas appears more relaxed than was apparent from the
ROSAT data.
\item The properties of this diffuse X-ray emission are subject
to large errors, but seem to fit rather well onto trends established
by other X-ray bright groups. HCG 16 appears to lie at the extreme 
poor end of the spectrum of X-ray emitting groups.
\end{itemize}
These properties suggest that HCG 16 is a genuinely bound
group, with a potential well in which gas has collected
and been heated, in a manner analogous to other X-ray bright groups.
In other words, the group has collapsed, even if we cannot state if it has 
already  virialized. A system which is dense in three-dimensions
is also strongly suggested by the extraordinary degree of activity
seen in its galaxies, which has presumably been triggered
by galaxy interactions.

However, HCG 16 is so far unique amongst groups with detectable
hot intergalactic gas, in containing no bright early-type galaxies.
One possible explanation, is that HCG 16 is a rare example of a group
which has collapsed in a single event (rather than building
through a sequence of mergers) very recently, so that the processes
which lead to the conversion of late-type into early-type galaxies
(mergers, interactions, stripping etc.) have not yet had time
to run their course. If this picture is correct, then the
implications are substantial. Typical X-ray bright groups, dominated
by a central bright early-type galaxy, represent only a minority of
groups. It is therefore very important to establish whether more
typical spiral-dominated groups have similar gas content.
If they do, then such groups may constitute the dominant component
of the baryonic content of the Universe (Fukugita et al 1998).
The properties of HCG 16 lend tentative support to this idea.

There is one fly in the ointment. If HCG 16 is a system currently
near maximum collapse, then its velocity dispersion should be
rather {\it high} (Mamon 1993). However, after correction
for measurement errors, the velocity dispersion of the group
is actually unusually {\it low}: 99~km~s$^{-1}$ if only the 
central four galaxies catalogued by Hickson are considered,
and 76~km~s$^{-1}$ if the additional three Ribeiro et al. (1996)
galaxies are included (Mamon, private communication). 
In terms of the parameter 
$\beta_{\rm spec} = (\mu m_p \sigma^2)/k T$, the ratio between
the specific energy on the galaxies, and that in the gas, which is
normally approximately unity in clusters, and rather lower in groups
(Helsdon \& Ponman 2000a), HCG 16 has the remarkably low value,
$\beta_{\rm spec} \sim 0.1$. This is difficult to understand, unless
we happen to be viewing HCG 16 at an angle such that most
of the galaxy motions lie within the plane of the sky.

In order to establish whether HCG 16 is truly a recently collapsed
system with a fortuitous alignment of galaxy orbits, or whether
we must seek some other explanation for its remarkable
combination of properties, we require deep X-ray studies of other
groups with similar optical characteristics. Such investigations
should be carried out with XMM and Chandra over the next few years.

\begin{acknowledgements}
We are grateful  to the referee, Gary Mamon, 
for interesting conversations on the status of HCG 16 and for the very 
useful comments of his referee report. We are grateful to Monique Arnaud
 for helpful discussions throughout this work and invaluable assistance in the estimation of the PSF. We also thank G.W. Pratt for contributions to the analysis and for very useful comments on the original  manuscript.
Our thanks to R. Gastaud and D.M. Neumann
for software support and to R. Rothenflug for help in the initial stages of the analysis.
This research has made use of the SIMBAD database,operated at CDS, Strasbourg, France. The paper is based on observations obtained with XMM-Newton, an ESA science mission with instruments and contributions directly funded by ESA Member States and the USA (NASA). EPIC was developed by the EPIC Consortium led by the Principal Investigator, Dr. M. J. L. Turner. The consortium comprises the
following Institutes: University of Leicester, University of
Birmingham, (UK); CEA/Saclay, IAS Orsay, CESR Toulouse, (France);
IAAP Tuebingen, MPE Garching,(Germany); IFC Milan, ITESRE Bologna,
IAUP Palermo, Italy. EPIC is funded by: PPARC, CEA,CNES, DLR and ASI.
\end{acknowledgements}

\end{document}